\documentclass[aps,prl,twocolumn,amsmath,longbibliography]{revtex4-2}  
\usepackage{bbm}
\usepackage{mathrsfs}
\usepackage{amsmath}
\usepackage{amsfonts}
\usepackage[colorlinks=true, 
            citecolor=blue, 
            anchorcolor=blue, 
            linkcolor=blue, 
            urlcolor=blue]{hyperref}
\usepackage{graphicx,epstopdf}
\usepackage{subfigure}
\usepackage{epsfig}
\usepackage{dcolumn}
\usepackage{bm}
\usepackage{color}
\usepackage{natbib}
\usepackage{amssymb} 
\usepackage{xcolor}
\usepackage{braket}
\usepackage{float}
\usepackage{balance}
\usepackage{afterpage}
\makeatletter
\renewcommand{\@fnsymbol}[1]{%
    \ensuremath{%
        \ifcase#1\or 
            \dagger
        \or \ddagger
        \or \S
        \or \P
        \or \|
        \or \dagger\dagger
        \or \ddagger\ddagger
        \or \S\S
        \or \P\P
        \or \|\|
        \else \@ctrerr \fi
    }%
}
\makeatother

\begin{document}
\title{Switchable axionic magnetoelectric effect via spin-flop transition in topological antiferromagnets}

\author{Yiliang Fan$^{1,\dagger}$, Rongxiang Zhu$^{1,\dagger}$, Tongshuai Zhu$^{3,4}$, Jianzhou Zhao$^{7}$, Huaiqiang Wang$^{2,6,\ast}$, and Haijun Zhang$^{1,5,6,\ast}$}

\affiliation{
 $^1$ National Laboratory of Solid State Microstructures, School of Physics, Nanjing University, Nanjing 210093, China\\
 $^2$ Center for Quantum Transport and Thermal Energy Science, Institute of Physics Frontiers and Interdisciplinary Sciences, School of Physics and Technology, Nanjing Normal University, Nanjing 210023, China\\
 $^3$ College of Science, China University of Petroleum (East China), Qingdao 266580, China\\
 $^4$ School of Materials Science and Engineering, China University of Petroleum (East China), Qingdao 266580, China\\
 $^5$ Jiangsu Key Laboratory of Quantum Information Science and Technology, Nanjing University, China\\
 $^6$ Jiangsu Physical Science Research Center, Nanjing 210093, China\\
 $^7$ Department of Physics, School of Science, Tianjin University, Tianjin 300354, China
}

\begin{abstract}
The MnBi$_2$Te$_4$ material family has emerged as a key platform for exploring magnetic topological phases, most notably exemplified by the experimental realization of the axion insulator state. While spin dynamics are known to significantly influence the axion state, a profound understanding of their interplay remains elusive. In this work, we employ an antiferromagnetic spin-chain model to demonstrate that an external magnetic field induces extrinsic perpendicular magnetic anisotropy. We find that an in-plane field stabilizes the antiferromagnetic order, whereas an out-of-plane field destabilizes it and triggers spin-flop transitions. Remarkably, near the surface spin-flop transition in even-layer MnBi$_2$Te$_4$ films, the axion insulator state undergoes a sharp switching behavior accompanied by distinct magnetoelectric responses. Furthermore, we propose that this switchable axionic magnetoelectric effect can be utilized to convert alternating magnetic field signals into measurable square-wave magneto-optical outputs, thereby realizing an axionic analog of a zero-crossing detector. Our findings could open a pathway toward potential applications of axion insulators in next-generation spintronic devices.

\end{abstract}

\thanks{These authors contributed equally to this work}
\email{zhanghj@nju.edu.cn}
\email{hqwang@njnu.edu.cn}

\maketitle

\begin{figure}[t]
    \centering
    \includegraphics[width=3.4in]{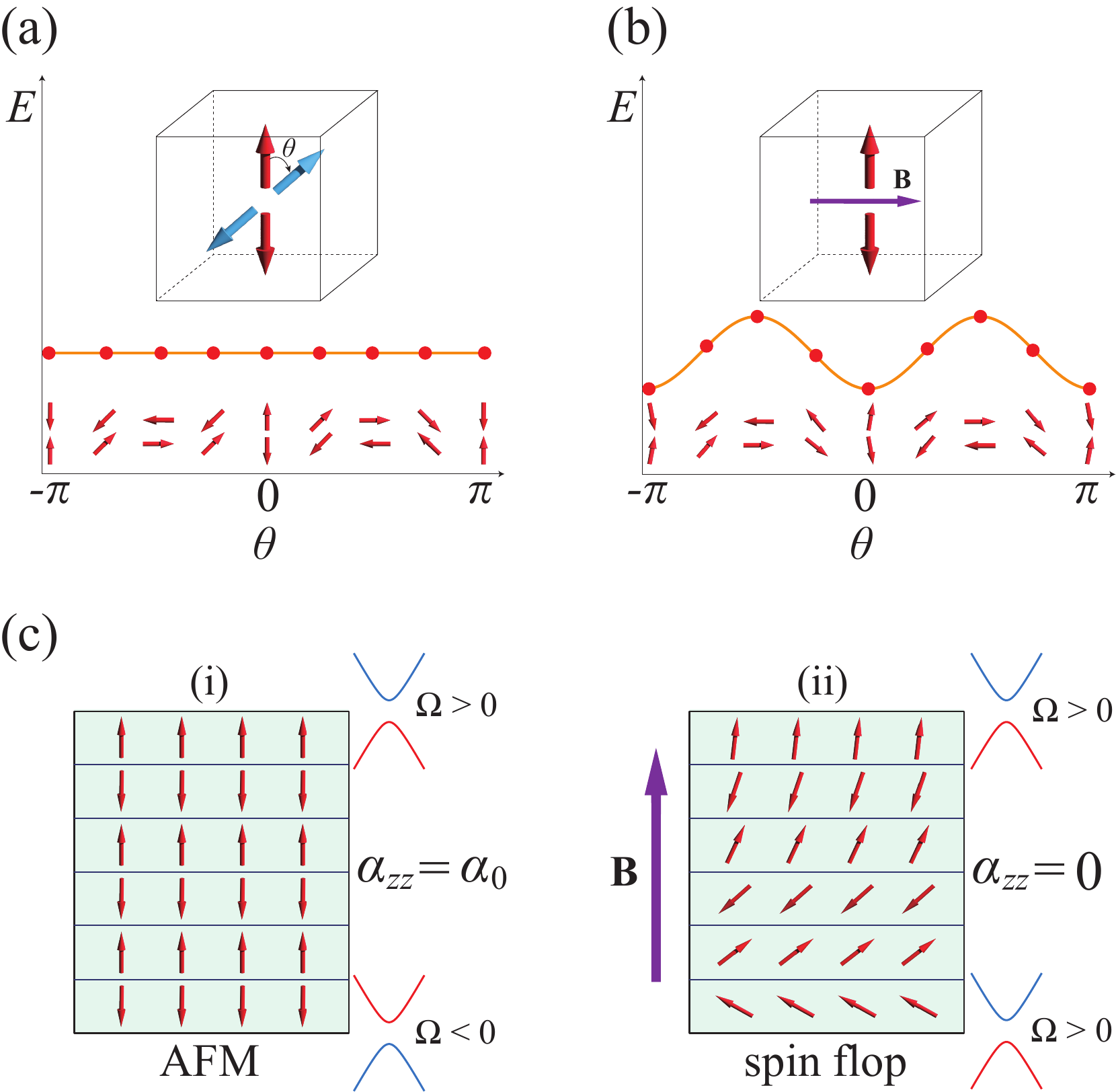}
    \caption{(a, b) Schematics of the field-induced perpendicular magnetic anisotropy. (a) In the absence of an external field, the magnetic energy landscape is isotropic with respect to the Néel vector orientation. (b) Under an applied magnetic field, two energy minima emerge where the Néel vector aligns perpendicular to the field direction. (c) Illustration of spin configurations, Berry curvatures of surface states, and out-of-plane magnetoelectric coefficient for even-SL AFM TIs (i) before and (ii) after the spin-flop transition induced by an external field.}
    \label{fig1}
\end{figure}
In topological field theory, the electromagnetic response of three-dimensional insulators is described by an axion action $S_{\theta}=(\theta/2\pi)(\alpha/2\pi)\int \mathrm{d}^3x\mathrm{d}t\mathbf{E}\cdot\mathbf{B}$~\cite{li_dynamical_2010,qi_rmp_2011}, where $\alpha=e^2/\hbar c$ is the fine-structure constant, $\mathbf{E}$ and $\mathbf{B}$ are the electric and magnetic fields, respectively, and $\theta$ is a dimensionless pseudoscalar field known as the axion field~\cite{peccei_mathrmcp_1977}. When the insulator preserves either time-reversal or spatial-inversion symmetry, $\theta$  becomes quantized to either 0 or $\pi$ (mod 2$\pi$), corresponding to topologically trivial and nontrivial phases, respectively. The quantized $\theta$ gives rise to a range of interesting phenomena, such as the topological magnetoelectric effect~\cite{qi_topological_2008,essin_magnetoelectric_2009,nomura_surface_2011,dziom_observation_2017}, quantized magneto-optical Faraday/Kerr rotation~\cite{maciejko_topological_2010,okada_terahertz_2016,wu_quantized_2016,ahn_theory_2022}, and the image magnetic monopole effect~\cite{qi_inducing_2009}. Realizing those phenomena requires gapped surface states by breaking the time-reversal symmetry (TRS) on the surface of topological insulators (TIs)~\cite{feng_observation_2015,mogi_magnetic_2017,xiao_realization_2018,nenno_axion_2020}. The recently discovered layered antiferromagnetic (AFM) TI MnBi$_2$Te$_4$~\cite{yan_gong_experimental_2019,zhang_topological_2019,otrokov_prediction_2019,li_intrinsic_2019,rienks_large_2019,chen_intrinsic_2019,chen_2019,otrokov_unique_2019,hao_gapless_2019,li_dirac_2019,zhu_floquet_2023,li_progress_2023} featuring both a nontrivial $\theta=\pi$ term and gapped surface states, has emerged as a prominent platform for topological magnetoelectric effects. For instance, odd-septuple-layer (SL) MnBi$_2$Te$_4$ films exhibit the quantum anomalous Hall (QAH) effect ~\cite{deng_quantum_2020,ge_high-chern-number_2020,lian_antiferromagnetic_2025,wang_towards_2025,zhang_zero-field_2025}, while even-SL films feature the axion insulator state~\cite{zhang_topological_2019,liu_robust_2020,zhu_tunable_2021,gao_layer_2021,lin_direct_2022,li_giant_2023,qiu_observation_2025} and quantum-metric-induced nonlinear transport~\cite{wang_quantum-metric_2023,gao_quantum_2023,li_quantum_2024,yan_unification_2024}.

Rich phenomena related to MnBi$_2$Te$_4$ stem from the interplay between its nontrivial band topology and distinctive $A$-type AFM order~\cite{xiao_nonlinear_2021,zhu_axionic_2022}, where each SL features out-of-plane ferromagnetic ordering and adjacent SLs are coupled antiferromagnetically. Interestingly, the AFM spin dynamics in layered MnBi$_2$Te$_4$ films are expected to give unconventional physical phenomena not observed in traditional magnetic materials~\cite{sass_robust_2020,li_competing_2020,yang_odd-even_2021,ovchinnikov_intertwined_2021,bac_topological_2022,lian_antiferromagnetic_2025}. For example, a cascade of quantum phase transitions induced by spin flips and flops has been discovered in the QAH state of MnBi$_2$Te$_4$~\cite{lian_antiferromagnetic_2025,wang_cascade_2025}. However, to our knowledge, the influence of AFM spin dynamics on the axion insulator state and the resultant topological magnetoelectric effect in MnBi$_2$Te$_4$-like layered systems yet remains unexplored.

In this work, based on an AFM spin-chain model for layered antiferromagnets, we demonstrate that an applied magnetic field effectively induces emergent magnetic anisotropy perpendicular to the field direction, as illustrated in Figs.~\ref{fig1}(a) and (b). Specifically, a magnetic field applied perpendicular to the easy axis enhances the magnetic anisotropy and stabilizes the AFM order, whereas a field parallel to the axis destabilizes the AFM order and further triggers a spin-flop transition. Notably, the spin-flop transition profoundly influences the axion insulator state in MnBi$_2$Te$_4$ films, thereby triggering a drastic change of the axionic magnetoelectric response, as shown by distinct surface Berry curvature configurations and out-of-plane magnetoelectric coefficients in Fig.~\ref{fig1}(c). Furthermore, we propose that near the spin-flop transition point, a weak out-of-plane AC magnetic field can periodically switch the axionic magnetoelectric response each time the field passes through zero. This conversion of the sinusoidal input field into a square-wave axionic response output can be experimentally read out via magneto-optical spectroscopy, thereby realizing an axionic analog of a zero-crossing detector common in signal processing. 

\emph{Field-induced perpendicular magnetic anisotropy and spin-flop transition}. We model the magnetic structure of even-SL MnBi$_2$Te$_4$ film exhibiting the axion insulator state as a one-dimensional AFM spin chain with $2N$ sites, representing each ferromagnetic SL by an effective spin ($\textbf{S}_i$) that is antiferromagnetically coupled to its nearest neighbors along the out-of-plane ($z$) direction. Under an external magnetic field $\bf{B}$, the spin-chain Hamiltonian can be described by
\begin{equation}
    \begin{aligned}
    H=&-g\mu_{\mathrm{B}}\sum^{2N}_{i=1}\mathbf{B}\cdot \mathbf{S}_i+J\sum_{i=1}^{2N-1}\mathbf{S}_i\cdot \mathbf{S}_{i+1}\\
&-D_z\sum_{i=1}^{2N}(S_i^z)^2+KN^2\left(\mathbf{B}\cdot\mathbf{L}\right)^2.
    \end{aligned}                                                               
\label{eq1}
\end{equation}
Here, the first term represents the conventional linear Zeeman coupling between each spin and magnetic field, with $g$ and $\mu_{\mathrm{B}}$ denoting the Land\'{e} $g$-factor and Bohr magneton, respectively. The second term describes the AFM exchange coupling ($J>0$) between neighboring SLs. The third term accounts for the intrinsic uniaxial magnetocrystalline anisotropy along the out-of-plane direction, characterized by the anisotropy constant $D_z$. The last term is a symmetry-allowed biquadratic higher-order coupling between the magnetic field and the N\'{e}el vector~\cite{khomskii_basic_2010} which is defined as $\mathbf{L}=(\sum_{i\in A}\mathbf{S}_i-\sum_{i\in B}\mathbf{S}_i)/N$ with $A$ and $B$ labelling the two AFM sublattices. It is noteworthy that this term, which has not been considered before, also makes a significant contribution to the magnetic anisotropy perpendicular to the applied field, as will be shown. The competition among these energy terms leads to a variety of intriguing magnetic phases and rich spin dynamics, most notably the spin-flop transition.

In the thermodynamic limit ($N\rightarrow\infty$), the system admits a macrospin description. Under this approximation, the sublattice magnetizations are represented by two macroscopic spins, $\mathbf{M}_{A,B} = g\mu_{\mathrm{B}}\sum_{i\in A,B}\mathbf{S}_i$, which can be parameterized as $\mathbf{M}_{A,B} \approx M(\sin\theta_{A,B}\hat{x} + \cos\theta_{A,B}\hat{z})$, with $M = g\mu_{\mathrm{B}} N S$. To track the orientation of the Néel vector under an external field, we introduce two collective variables. The tilting angle $\theta=(\theta_{A}+\theta_{B}-\pi)/2$ defines the orientation of the Néel vector relative to the $z$-axis, while $\theta_c=(\theta_{A}-\theta_{B})/2$ characterizes the canting angle between the two sublattice spins. Then the average magnetic energy $E(\theta,\theta_c)$ can be derived [see the Supplemental Material (SM)~\cite{SM} for details]. Following the principle of energy minimization, the effective energy $E(\theta)$ as a function of the tilting angle $\theta$ is obtained by minimizing the total energy $E(\theta,\theta_c)$ with respect to the canting angle $\theta_c$.

\begin{figure}[htbp] 
    \centering
    \includegraphics[width=3.4in]{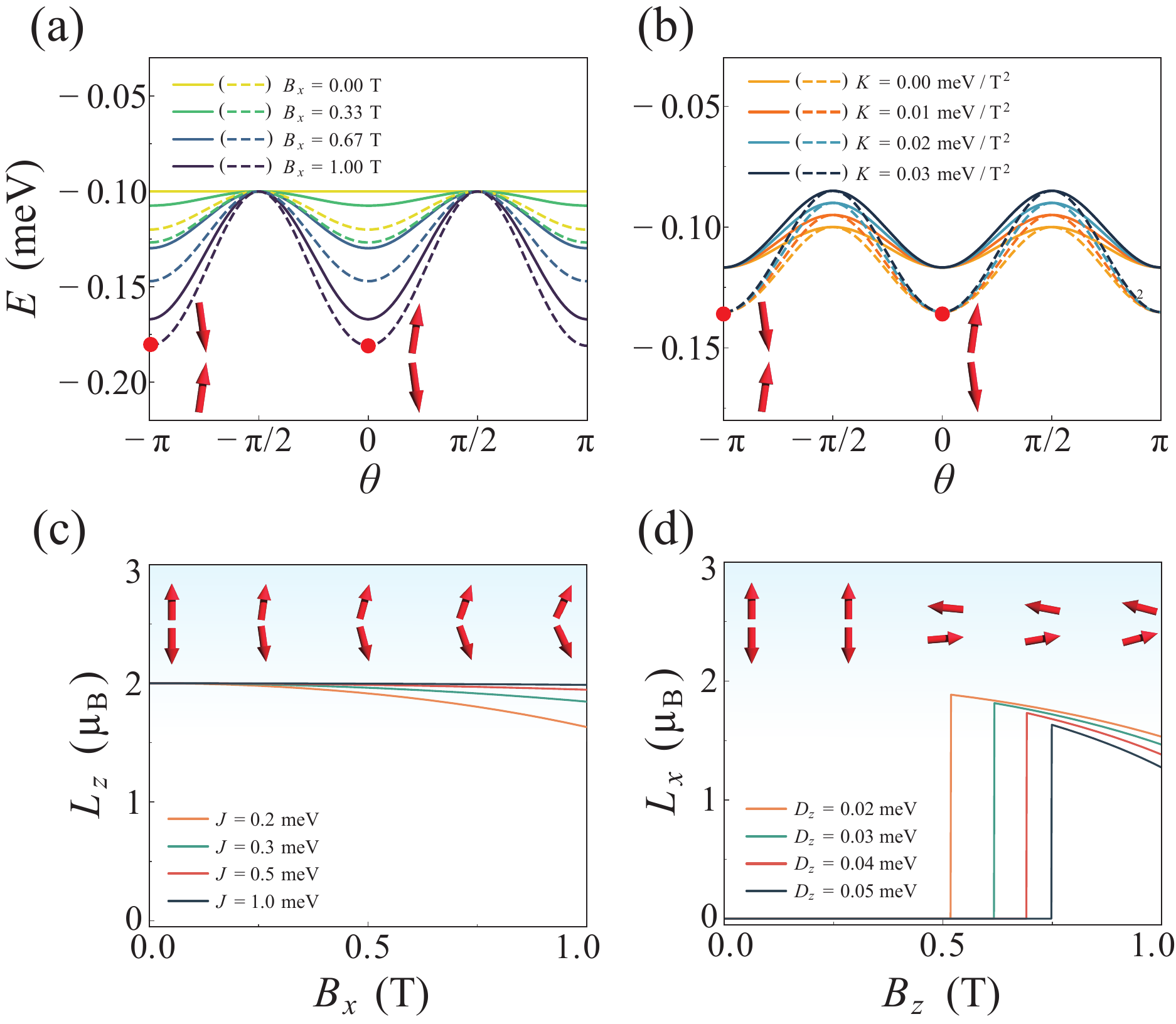}
    \caption{(a) Magnetic anisotropy energy versus N\'{e}el  vector orientation under in-plane magnetic fields ($B_x$), considering only the linear Zeeman coupling: without the uniaxial anisotropy (solid lines) and with out-of-plane anisotropy (dashed lines, $D_z=0.02$ meV). (b) Magnetic energy, including both linear and higher-order coupling between the Néel vector and magnetic field, plotted as a function of higher-order coupling strength at fixed $B_x=0.5$ T: without (solid lines) and with out-of-plane uniaxial anisotropy (dashed lines). (c) Out-of-plane component ($L_z$) of the N\'{e}el vector as a function of in-plane field ($B_x$) for different interlayer exchange couplings ($J$). Insets: Schematics of spin configurations under different $B_x$ with $J$ = 0.2 meV. (d) In-plane component ($L_x$) of the N\'{e}el vector as a function of out-of-plane field $B_z$ for different out-of-plane anisotropic strengths. Insets: Schematics of spin configurations under different $B_z$ with $D_z$ = 0.02 meV.}
    \label{fig2}
\end{figure}

We begin by considering a simplified case containing only the isotropic AFM exchange coupling and the linear Zeeman term, with the magnetic field applied along the $x$-direction. The solid energy curves $E(\theta)$ in Fig.~\ref{fig2}(a) show that the Zeeman term breaks the isotropy of the AFM structure. This symmetry breaking generates a pair of degenerate minima at $\theta$ = 0 and $\pi$, corresponding to the N\'{e}el vector aligning along the $+z$ and $-z$ directions, respectively. Thus, the applied magnetic field establishes an extrinsic magnetic anisotropy, thereby pinning the easy axis perpendicular to itself. This conclusion is further confirmed by the field dependence of the N\'{e}el vector component $L_z$ shown in Fig.~\ref{fig2}(c). Upon applying an infinitesimal $B_x$, $L_z$  immediately jumps to a finite value of $2\mu_B$, signaling the abrupt alignment of the N\'{e}el vector along the easy axis. Despite the increasing spin canting along the $x$-direction with $B_x$, the Néel vector remains robustly pinned along the $z$-axis. Furthermore, the higher-order coupling term enhances the perpendicular magnetic anisotropy by contributing a positive energy cost for any misalignment of the Néel vector from the direction normal to the field [solid lines in Fig.~\ref{fig2}(b)].

We now proceed to incorporate the out-of-plane uniaxial magnetic anisotropy term inherent to layered MnBi$_2$Te$_4$ films and investigate two representative configurations with in-plane ($B_x$) and out-of-plane ($B_z$) magnetic fields, respectively. Under an in-plane field $B_x$, the Zeeman and higher-order coupling terms both reinforce the perpendicular orientation of the N\'{e}el vector, thereby cooperating with the intrinsic uniaxial anisotropy to strengthen the out-of-plane easy axis, as evidenced by the dashed lines in Figs.~\ref{fig2}(a) and~\ref{fig2}(b). In contrast, an out-of-plane field $B_z$ favors an in-plane N\'{e}el vector, which competes with the intrinsic out-of-plane anisotropy. Consequently, upon increasing $B_z$ to a critical value, the system undergoes an AFM spin-flop transition, characterized by an abrupt reorientation of the Néel vector from out-of-plane to in-plane directions [Fig.~\ref{fig2}(d)]. Furthermore, we have performed atomistic spin simulations based on the Landau-Lifshitz-Gilbert equation (see SM~\cite{SM}), which are in good agreement with our theoretical analysis.

\begin{figure}[htbp]
	\centering
	\includegraphics[width=3.4in]{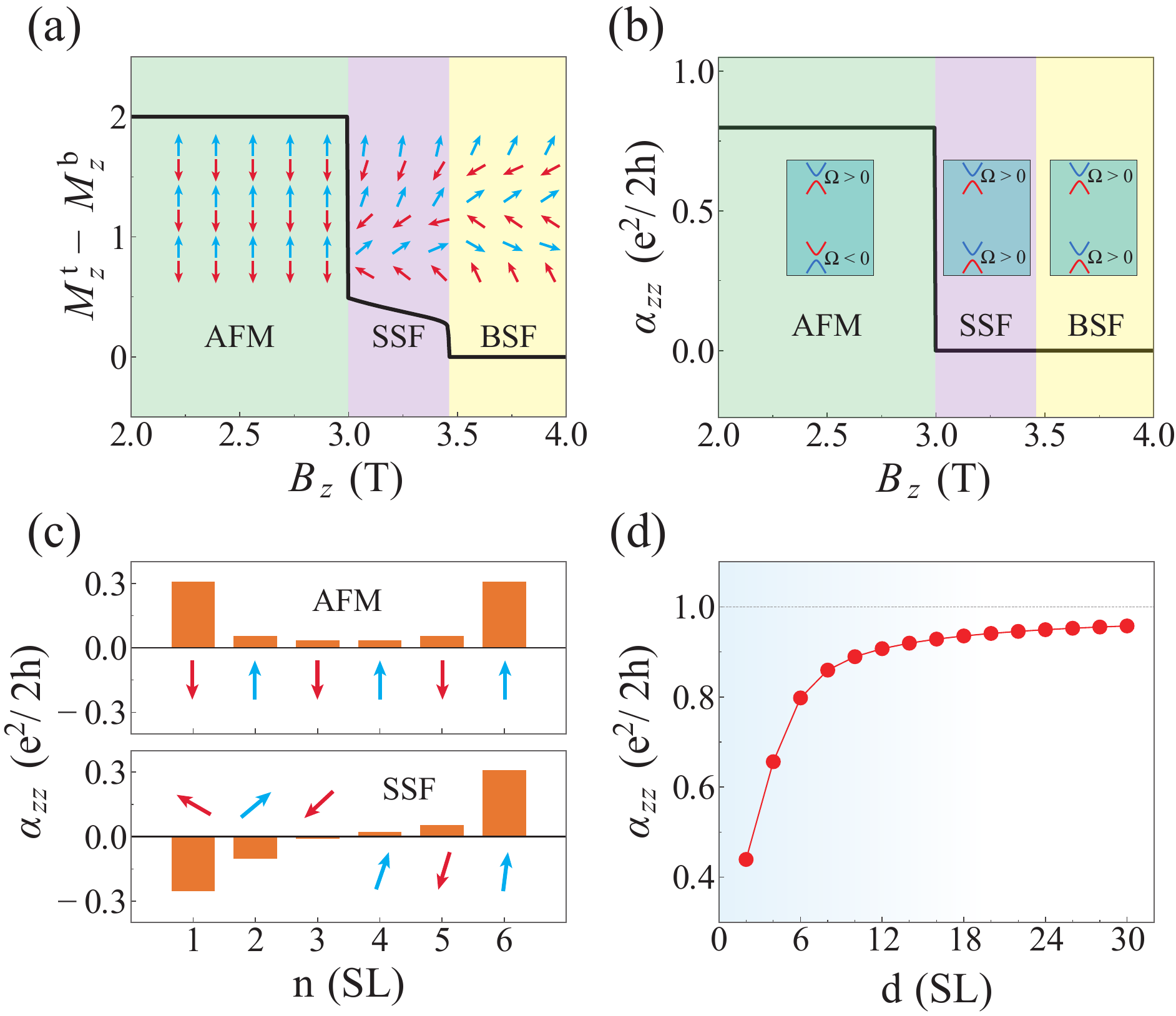}
	\caption{(a) Evolution of spin configurations in each SL and the difference between the $z$-component magnetic moments of top and bottom surfaces versus out-plane magnetic field $B_z$ in a six-SL MnBi$_2$Te$_4$ film, obtained from atomistic spin simulations. Surface spin-flop (SSF) and bulk spin-flop (BSF) transitions occur at $B_z=3.00$ T and $3.45$ T, respectively. (b) The out-of-plane magnetoelectric coupling coefficient $\alpha_{zz}$ and corresponding surface Berry curvature configurations as a function of out-of-plane magnetic field $B_z$. (c) Layer resolved magnetoelectric coupling coefficient $\alpha_{zz}$ in the (upper panel) AFM state and (lower panel) SSF state. (d) Thickness dependence of $\alpha_{zz}$, illustrating the finite-size effect. Detailed parameters in atomistic spin simulations and numerical calculations are provided in the SM~\cite{SM}.}
	\label{fig3}
\end{figure}

\emph{Axionic magnetoelectric response across the spin-flop transition.}
Given that the axion field in even-SL MnBi$_2$Te$_4$ films depends on the magnetic configurations~\cite{wang_dynamical_2020, liu_anisotropic_2020,zhu_tunable_2021}, the AFM spin-flop transition involving a drastic reconfiguration of spins is therefore expected to profoundly influence the axionic magnetoelectric response. Owing to the experimental availability of high-quality samples and advanced magnetoelectric measurement capabilities~\cite{qiu_observation_2025}, we select the six-SL MnBi$_2$Te$_4$ film as a representative system for our study. We model the spin-flop transition in a six-SL MnBi$_2$Te$_4$ film using atomistic spin simulations of a six-spin chain under out-of-plane magnetic fields, with input parameters obtained directly from first-principles calculations (see SM~\cite{SM} for details). Our simulations reveal two spin-flop transitions at critical fields of 3.00 T and 3.45 T [see Fig.~\ref{fig3}(a)], corresponding to the surface spin-flop (SSF) and bulk spin-flop (BSF) transitions, respectively. Before the SSF transition, the system retains the $A$-type AFM order. Upon the SSF transition, the system exhibits an asymmetrically canted AFM order. Specifically, the spins near the top surface remain close to the out-of-plane direction, while those at the bottom surface tilt significantly toward the in-plane direction, as evidenced by the difference in the $z$-components of their magnetic moments in Fig.~\ref{fig3}(a). This behavior originates from the interplay between the modified exchange couplings and magnetocrystalline anisotropy near the surface~\cite{yang_odd-even_2021,sass_robust_2020,robler_2004}. Beyond the BSF transition, the top and bottom surface spins cant at an identical angle, which is smaller than the uniform canting angle exhibited by bulk spins.

To investigate the axionic magnetoelectric response of MnBi$_2$Te$_4$ films, we employ the Dirac-fermion approach for layered magnetic TIs, where each SL is effectively modeled by two Dirac-cone states located on its top and bottom surfaces, respectively. Within this framework, an $N$-SL MnBi$_2$Te$_4$ can be described by the following Hamiltonian~\cite{lei_magnetized_2020,wang_dirac_2023,wang_three_2023} 
\begin{equation}
\begin{aligned}
	H = \sum^{N}_{\mathbf{k},ij=1} \Bigl[ & \bigl( v_F\tau_z(k_x\sigma_y - k_y\sigma_x) + \mathbf{m}_i\cdot\mathbf{\sigma} + \Delta_S\tau_x \bigr) \delta_{i,j} \\
	& + \Delta_D\tau_+ \delta_{j,i+1} + \Delta_D\tau_- \delta_{j,i-1} \Bigr] c_{\mathbf{k} i}^\dagger c_{\mathbf{k} j}.
\end{aligned}
\label{eq4}
\end{equation}
Here, $v_F$ is the Fermi velocity, $\sigma_i$ and $\tau_i$ are Pauli matrices acting in the spin and top-bottom surface subspaces, respectively, with $\tau_{\pm}=\tau_x\pm i\tau_y$. The $\mathbf{m}_i\cdot\mathbf{\sigma}$ term denotes the Zeeman coupling of each Dirac fermion to the effective exchange field from surrounding magnetic moments. The $\Delta_S$ and $\Delta_D$ terms represent the coupling of top and bottom Dirac surface states within the same SL and between neighboring SLs, respectively.

Based on the above Dirac-fermion model, we then use the Kubo formula (see SM for details)~\cite{zhu_tunable_2021,mei_2024,SM} to calculate the field-dependent out-of-plane axionic magnetoelectric response coefficient $\alpha_{zz}$, as shown in Fig.~\ref{fig3}(b). Before the SSF transition, $\alpha_{zz}$ of the six-SL MnBi$_2$Te$_4$ film exhibits a nonzero value of approximately 0.8 $e^2/2h$. Note that the reduction from the quantized value of $e^2/2h$ for an ideal axion insulator originates from the finite-size-induced hybridization between the top and bottom surface states~\cite{zhu_tunable_2021, liu_anisotropic_2020, fan_symmetry_2024}. This interpretation is confirmed by our thickness-dependent calculation of $\alpha_{zz}$ [Fig.~\ref{fig3}(d)], which shows a monotonic increase with film thickness and asymptotically approaches $e^2/2h$ in the thick limit. Upon the SSF transition, the coefficient $\alpha_{zz}$ abruptly drops to nearly zero and stays vanishingly small through the subsequent BSF transition. The drop of $e^2/2h$ across the SSF transition is caused by a band inversion and the associated Berry curvature reversal of the bottom surface state. This reversal occurs because the SSF reverses the $z$-component of the magnetic moment of the bottom SL, which inverts the sign of the TRS-breaking mass term for the surface state. This picture is supported by the distinct layer-resolved contributions to $\alpha_{zz}$ in the AFM and SSF states, as visible from the upper and lower panels of Fig.~\ref{fig3}(c), respectively.
\begin{figure}[htbp]
	\centering
	\includegraphics[width=3.4in]{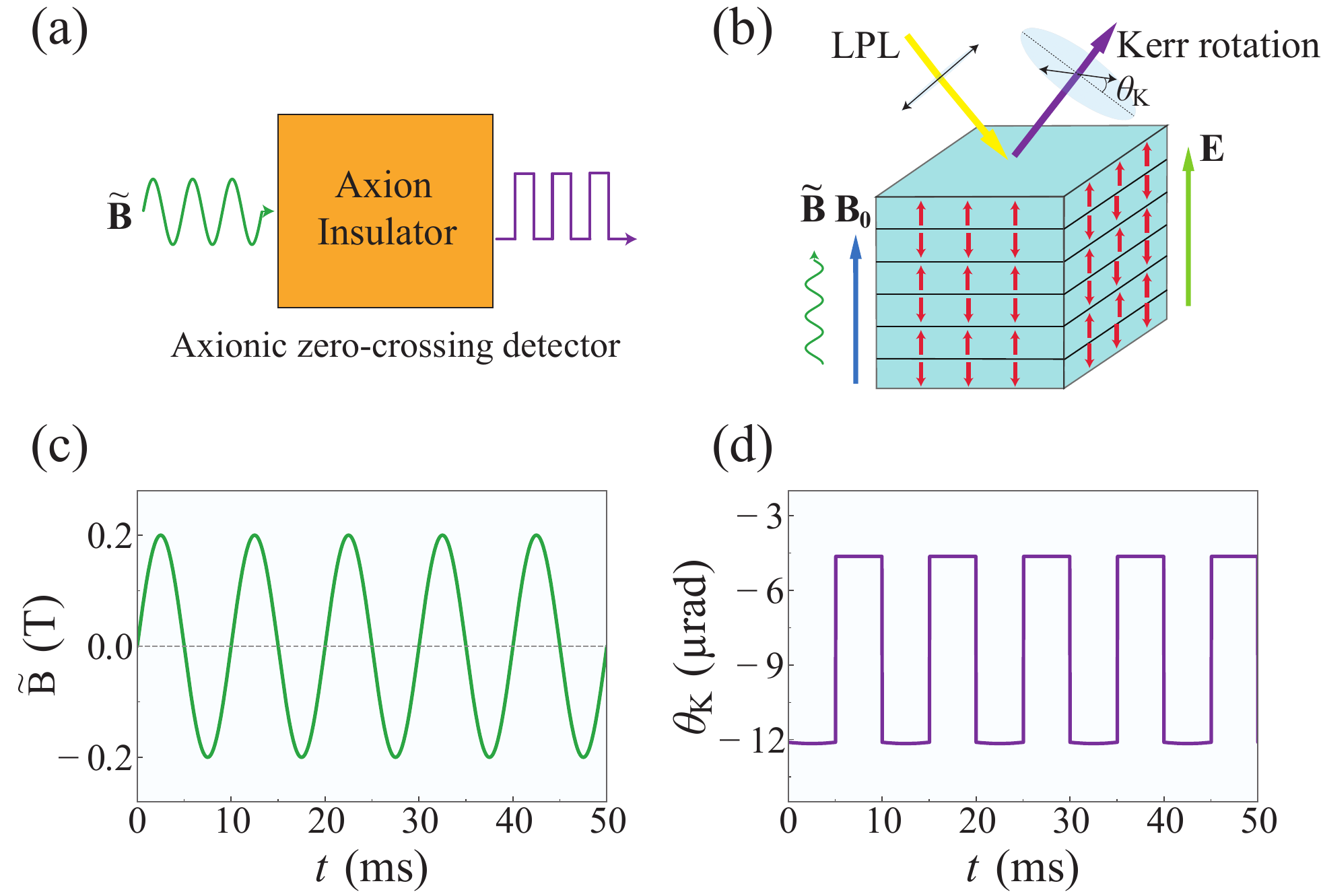}
	\caption{(a,b) Schematic of an axionic zero-crossing detector constructed from an even-SL MnBi$_2$Te$_4$ film, which transforms alternating input magnetic signals to square-wave outputs (a) and experimental setup (b), where a static out-of-plane magnetic field is applied to position the MnBi$_2$Te$_4$ film at its SSF transition point and an alternating out-of-plane magnetic field is further applied as the input signal. The output signal is obtained from the magneto-optical Kerr rotation angle induced by an out-of-plane electric field for an incident linearly polarized THz light. (c,d) The time-varying input alternating magnetic field (c) and the obtained square-wave-like output (d) from the corresponding Kerr angles under $48.4$-THz light. Detailed parameters in numerical calculations are provided in the SM~\cite{SM}.}
	\label{fig4}
\end{figure}

\emph{Axionic zero-crossing detector.} The switchable sharp change in the precedent axionic magnetoelectric coefficient $\alpha_{zz}$ at the SSF transition renders it highly sensitive to small deviations of the external magnetic field from the critical SSF transition field. This unique property can be harnessed to realize an axionic zero-crossing detector, as illustrated in Fig.~\ref{fig4}(a), which transforms continuous magnetic field signals to square waves. A zero-crossing detector is an electronic circuit whose core functionality is to detect when an input signal passes through zero potential (the "zero-crossing point") and produce an instantaneous switch in the output signal at that precise moment. We design the axionic zero-crossing detector as follows. An out-of-plane static magnetic field is applied to position the six-SL MnBi$_2$Te$_4$ film near its SSF transition point, while a target alternating field is applied along the same direction to serve as the input signal [Fig.~\ref{fig4}(b)]. The time-varying target magnetic field drives rapid switching of $\alpha_{zz}$ between its two discrete values, which could be measured by magneto-optical responses acting as square-wave output signals.

For a realistic estimate, the static critical SSF field and alternating target field are set to $B_0$ = 3.0 T and  $\tilde{B}$ = 0.2 $\sin(2\pi\nu t)$ T, respectively, with $\nu$ = 100 kHz~\cite{p_kollar_ac_2010}. For magneto-optical readout, we apply a static out-of-plane electric field $E_z$ ($\epsilon E_z=0.5$ V/nm, where $\epsilon$ is the permittivity)~\cite{qiu_observation_2025} and measure the Kerr rotation angle of incident linearly polarized THz light. Theoretical predictions of Kerr rotation angles for different photon energies and magnetic configurations can be implemented within the Dirac-fermion framework. The electric field is incorporated as a layer-resolved potential term in the Dirac-fermion model, which is determined self-consistently by solving the discretized Poisson equation with parameters obtained directly from first-principles calculations of six-SL MnBi$_2$Te$_4$~\cite{SM}. Figure~\ref{fig4}(d) presents the resulting time-varying Kerr rotation angle under $48.4$-THz ($\sim 0.2$ eV) light, which transforms the sinusoidal input field [Fig.~\ref{fig4}(c)] into square-wave magneto-optical outputs, thus realizing the zero-crossing detection functionality. It is also worth mentioning that the predicted Kerr angles are indeed experimentally accessible (-12$\sim$-4.5 $\mathrm{\mu rad}$), as demonstrated in recent measurements on MnBi$_2$Te$_4$ films~\cite{qiu_observation_2025}.

\emph{Summary.} We present a comprehensive study of the field-tunable axion electrodynamics in MnBi$_2$Te$_4$ films. Using the spin-chain model for layered antiferromagnets, we reveal that a magnetic field induces emergent perpendicular magnetic anisotropy, which can either stabilize the AFM order or trigger a spin-flop transition. This spin-flop transition profoundly modifies the axion insulator state and switches its magnetoelectric response. We leverage this sharp switching to propose and demonstrate the principle of an axionic zero-crossing detector, which transduces an AC magnetic field into a square-wave magneto-optical signal. These results pave the way for axion insulators in next-generation spintronic devices.

\emph{Acknowledgements.}
The work is supported by the National Key Research and Development Program of China (Grants No. 2024YFA1409100 and No. 2021YFA1400400); the Natural Science Foundation of Jiangsu Province (Grants No. BK20233001, No. BK20243011 and No. BK20252117), the Natural Science Foundation of China (Grant No. 12534007 and No. 92365203), the e-Science Center of Collaborative Innovation Center of Advanced Microstructures, the Postdoctoral Fellowship Program of CPSF (Grant No. GZC20242012), Shandong Provincial Natural Science Foundation (Grant No. ZR2024QA095) and the Fundamental Research Funds for the central Universities (Grant No. 23CX06063A).

\bibliography{ref}

\end{document}